\documentclass[preprint,10pt]{elsarticle}

\makeatletter
\def\ps@pprintTitle{%
 \let\@oddhead\@empty
 \let\@evenhead\@empty
 \def\@oddfoot{}%
 \let\@evenfoot\@oddfoot}
\makeatother

\usepackage{amssymb}

\usepackage[utf8]{inputenc}
\usepackage{amsmath}
\usepackage{amsfonts}
\usepackage{enumitem}  
\usepackage[ruled,vlined,linesnumbered,resetcount]{algorithm2e}
\usepackage{tikz}
\usetikzlibrary{positioning}
\usepackage{caption}
\usepackage{subcaption}

\usepackage{xcolor}
\usepackage{graphicx}
\usepackage[normalem]{ulem}

\usepackage{bm}

\usetikzlibrary{arrows.meta}

\usepackage{minibox}
\usepackage[normalem]{ulem}

\newcommand{\myfrac}[2]{\displaystyle{\frac{#1}{#2}}}
\newcommand{\somme}[3]{\displaystyle{\overset{#3}{\underset{#1=#2}{\sum}}}}

\newcommand{\deriveepartielle}[2]{\displaystyle{\partial_{#2} #1}}

\newcommand{\vect}[1]{ {\bm{#1}}}

\newcommand{\ctrlvar}{\delta}

\newcommand{\statevecU}{\vect{v}}

\newcommand{\Sbase}[1]{\Phi_{#1}}

\newcommand{\Tbase}[1]{{\Lambda_{{{#1}}}}}

\newcommand{\nbrModes}{q}

\newcommand{\nbrParam}{N_{p}}
\newcommand{\nbrSnap}{N_{s}}
\newcommand{\nbrBases}{N_{p}}
\newcommand{\dimVecorsSnap}{N_x}

\newcommand{\trainedSnapMat}[1]{\bm{\mcal{Y}}_{_{{_{#1}}}}}
\newcommand{\trainedredSnapMat}[1]{\mcal{Y}_{_{{#1}}}}
\newcommand{\untrainedredSnapMat}{\tilde{\mcal{Y}}}

\newcommand{\mcal}[1]{\mathcal{#1}}
\newcommand{\mbb}[1]{\mathbb{#1}}

\newcommand{\txt}[1]{\textnormal{#1}}

\newcommand{\constraint}[0]{\mathcal{N}}

\newcommand{\functional}[0]{\mathcal{J}}

\newcommand{\Mat}[1]{\bm{#1}}

\newcommand{\trainingparam}[1]{\ctrlvar_{#1}}

\usepackage[a4paper, total={5.5in, 9.in}]{geometry}

\usepackage{floatrow}
\newfloatcommand{capbtabbox}{table}[][\FBwidth]

\SetKwInput{KwOffline}{Offline}
\SetKwInput{KwOnline}{Online}
\usepackage[export]{adjustbox}
\begin{document}

\begin{frontmatter}



\author{M. Oulghelou\fnref{label1}}
\fntext[label1]{mourad.oulghelou@univ-lr.fr}
\author{C. Beghein\fnref{label2}}
\fntext[label2]{cbeghein@univ-lr.fr}
\author{C. Allery\fnref{label3}}
\fntext[label3]{cyrille.allery@univ-lr.fr}

\address{LaSIE, UMR-7356-CNRS, Universit\'e de La Rochelle P\^ole Science et Technologie,
Avenue Michel Cr\'epeau, 17042 La Rochelle Cedex 1, France.}



\title{Data-Driven Optimization Approach for Inverse Problems : Application to Turbulent Mixed-Convection Flows}


\begin{abstract}
Optimal control of turbulent mixed-convection flows has attracted considerable attention from researchers. Numerical algorithms such as Genetic Algorithms (GAs) are powerful tools that allow to perform global optimization. These algorithms are particularly of great interest in complex optimization problems where cost functionals may lack smoothness and regularity. In turbulent flow optimization, the hybridization of GA with high fidelity Computational Fluid Dynamics (CFD) is extremely demanding in terms of computational time and memory storage. Thus, alternative approaches aiming to alleviate these requirements are of great interest. Nowadays, data driven approaches gained attention due to their potential in predicting flow solutions based only on preexisting data. In the present paper, we propose a near-real time data-driven genetic algorithm (DDGA) for inverse parameter identification problems involving turbulent flows. In this optimization framework, the parametrized flow data are used in their reduced form obtained by the POD (Proper Orthogonal Decomposition) and solutions prediction is made by interpolating the temporal and the spatial POD subspaces through a recently developed Riemannian barycentric interpolation. The validation of the proposed optimization approach is carried out in the parameter identification problem of the turbulent mixed-convection flow in a cavity. The objective is to determine the inflow temperature and inflow velocity corresponding to a given temperature distribution in a restricted area of the spatial domain. The results show that the proposed genetic programming optimization framework is able to deliver good approximations of the optimal solutions within less than two minutes.
\end{abstract}


\end{frontmatter}
\newpage
\section{Introduction}
Decreasing energy consumption of buildings is an important aspect of the reducing of global warming. 
However, the energy reduction has to be compromised with the quality of thermal comfort inside buildings. To achieve that, optimization applied to indoor airflows, which is aimed at determining optimal flow values for some well chosen parameters are of great interest. 
The optimization objective can be expressed in the whole or a part of the domain, in terms of field variables such as inlet velocity, wall temperature, heat source, etc. For flows in buildings, which are mostly mixed convection turbulent flows, high fidelity solvers are privileged for parameter identification problems. 
A usual class of flow optimization algorithms consists in standard gradient descent algorithms using high fidelity adjoint equations. The search direction is computed as the functional cost sensitivity over the design variables and the solution is moved along until an optimal solution is reached. This approach was used for instance by Liu et al. to find optimal thermo-fluid boundary conditions in a two-dimensional cavity \cite{Liu2015} and to optimize the air supply location, size, and parameters in a two dimensional non isothermal ventilated cavity \cite{Liu2016}. It was also used to optimize buoyancy-driven ventilation flows governed by Boussinesq equations \cite{NABI2017342,NABI2019104313}.
A major limitation of high fidelity adjoint-based algorithms is that they are more likely to stuck in local optima. To overcome this issue, a global optimization search can be carried out by Genetic Algorithms (GAs) \cite{Holland1975}. In the context of mixed-convection flows, high fidelity solvers combined with GA have been investigated and validated in \cite{XUE201377, DIAS2006}. Compared to high fidelity adjoint based optimization approach, high fidelity based GA is more efficient in terms of finding global optimal solutions, yet it requires a tremendous computing load, leading to turn the attention to techniques of model reduction. 
\vspace*{0.2cm}
\\
Reduced-order models have been extensively used in fluid dynamics in order to reduce the computational burden in optimization and control applications. Recently, POD/Galerkin reduced order models were successfully combined with optimization approaches allowing a drastic alleviation of the optimization computational effort. A standard approach developed by Tallet et al. \cite{Tallet_Elsevier} and Bergmann et al. \cite{BergmannCordier} consists in using high fidelity simulations to extract a POD basis representing the main structures of a set of snapshots sampled at different parameter values. The temporal dynamics is afterwards calculated by solving an ordinary system of differential equations resulting from Galerkin projection of the governing equations onto the global POD basis. By considering the global POD/Galerkin ROM as the state equations, a reduced scale optimization problem can be formulated and solved in near-real time. However, in many physical cases, the global POD/Galerkin ROM may experience issues of accuracy due to the overload of information in the global POD basis. Sophisticated subspace interpolation techniques such as the ITSGM (Interpolation on the Tangent Space of the Grassmann Manifold) proposed by Amsallem et al. \cite{Amsallem} is an efficient local method meant to restrict the ROM predictions to the wanted physical regime. In the context of the adjoint-based optimal control, the ITSGM/Galerkin ROM was successfully embedded in a suboptimal control strategy to achieve a near-real time optimal control of transfer phenomena \cite{OULGHELOUAMC2018}.
\vspace*{0.2cm}
\\
In the last two decades, interest in data driven model reduction techniques for flow problems is increasingly growing. Interestingly, the power of these methods is their dispense on the underlying mathematical model. Instead, they explore and learn the dynamics from preexisting data and deliver approximations that are expected to capture most of the dynamics of the flow. Numerous attempts have been carried out in this subject. Namely, one can refer to \cite{oulghelou2020} where a modified version of the ITSGM referred to as Bi-CITSGM designed for non-linear data interpolation is proposed, and to \cite{Cheng2020, Xiao2019323, San2019271, Yu20195896, Ahmed2019} where Artificial Neural Networks (ANN) were used with model reduction for the prediction of flow solutions. In the context of data driven optimization, the Bi-CITSGM has been used successfully in conjunction with GA  to control the flow past a circular cylinder and the flow in a lid driven cavity \cite{oulghelou2020}. In the same spirit, ANN are used in conjunction with micro genetic algorithm (MGAs) for the optimization of the location of multiple discrete heat sources in a ventilated cavity \cite{MADADI2008}. In both cases, the Bi-ITSGM or ANN combined with GA demonstrated their ability to reach good suboptimal solutions within a near-real computational time.
%
%
%
%
\vspace*{0.2cm}
\\
In this paper, we formulate a new Data Driven Genetic Algorithm (DDGA) based on the Riemannian Barycentric interpolation of subspaces. This interpolation method is based upon the geometry of the manifold of fixed rank matrices studied in details in \cite{Massart2019}. It was initially used to interpolate low-rank solutions of the Luyapunov equations  resulting  from parametric linear input-output reduced order system \cite{Massar2020}, and recently adapted to interpolate the parametric Navier-Stokes Galerkin/ROM \cite{OulghelouSPsDROMArxiv}. In contrast to the Bi-CITSGM which needs a calibration phase for the interpolated POD subspaces, the barycentric interpolation naturally results in modes that are arranged according the POD energetic content. This property allows to interpolate the time and space quantities separately and eventually form the set of untrained solutions by simply combining them. 
The aim of the following study is to use a preexisting flow database to solve the inverse parameter identification problem involving the turbulent mixed-convection flow in a cavity. The optimization objective is to determine the inlet velocity and temperature that optimize the cost functional related to maintaining a desired temperature distribution inside a part of the spatial domain.
%
%
%
\vspace*{0.2cm}
\\
The remainder of this article is organized as follows: First, the studied Mixed convection inverse problem is presented in section 2. In section 3, the barycentric interpolation used for nonlinear parametrized data prediction is detailed. Next, the  proposed data driven Genetic Algorithm is outlined in section 4. In section 5, numerical experiments assessing the potential of this approach are carried out on the inverse problem involving the turbulent mixed-convection flow in a cavity. Finally, conclusions are drawn in section 6.

\section{Mixed convection inverse problem}
	\subsection{Optimization problem settings}
\begin{figure}[hbtp!]
\hspace*{-3cm}
\includegraphics[width=0.8\linewidth]{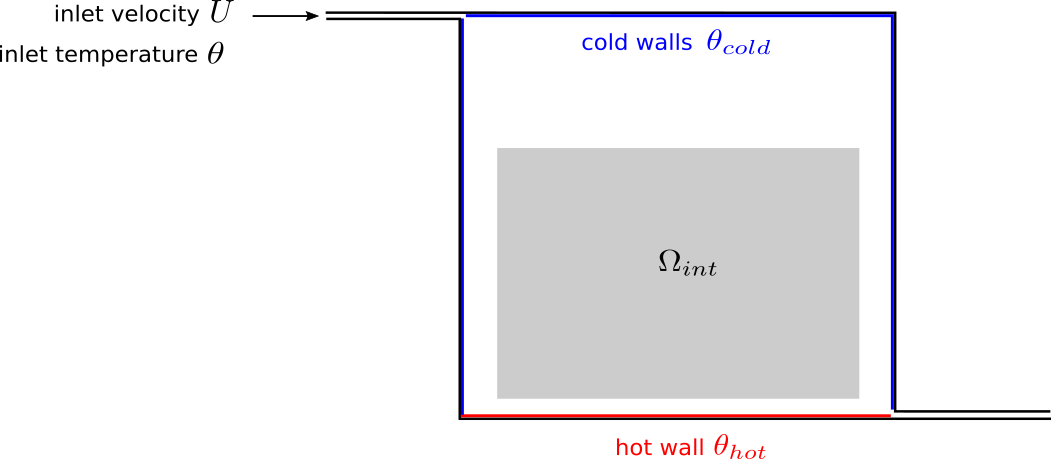}
\caption{Description of the studied mixed-convection flow}
\label{fig.Temperature_img}
\end{figure}

	This study focuses on the inverse problem of temperature distribution in a two-dimensional ventilated cavity, whose dimensions
are $1.04m\times1.04m $, and which is shown in figure \ref{fig.Temperature_img}.
The temperature $\theta_{hot}$ of the bottom wall of the cavity is higher than the temperature $\theta_{cold}$ of the other walls:
\begin{equation}
 \theta_{hot}=35^oC \text{ and } \theta_{cold}=15^oC
\end{equation}
The air inlet (resp. outlet) is located at the top left (resp. bottom
right) corner of the cavity. The vertical dimension of the air inlet (resp. outlet) is $0.018$ m (resp. $0.024$ m). At the inlet, the air velocity is denoted $U$, and the air temperature is $\theta$. The turbulent air flow in the
cavity is governed by the equations of mass conservation, momentum conservation, energy conservation, of an incompressible Newtonian
fluid with Boussinesq's assumption
\begin{equation}\label{EQ.Navier_Stokes}
\begin{cases}
\nabla\cdot\statevecU = 0 
\\
\rho \deriveepartielle{\statevecU}{t} + \rho \statevecU\cdot\nabla \statevecU  =  - \nabla p + \mu \Delta \statevecU + \rho \, \mathbf{g} \, \beta (\Theta - \Theta_{0})\bm{e}_y  + \nabla \bm{\sigma_t}
\\
\rho \,  c_p \,  \deriveepartielle{\Theta}{t}+ \rho \,  c_p \,  \statevecU\cdot\nabla \Theta = \lambda \,  \Delta \Theta + \nabla \mathbf{q_t}
\end{cases}
\end{equation}
where $\statevecU$, $\Theta$, $p$ are the time averaged velocity\footnote{In these equations, $\statevecU$, $\Theta$ and $p$ should have been written $\overline{\statevecU}$, $\overline{\Theta}$ and $\overline{p}$. To alleviate the notations in the reminder of the paper, the time averaged notation will not be used.}, temperature and pressure obtained with an Unsteady Reynolds Averaged Navier-Stokes (URANS)
turbulence model. $\rho$, $\mu$, $C_p$, $\lambda$ are the density, dynamic
viscosity, heat capacity and heat conductivity of the fluid at the reference temperature $\Theta_0$, $\mathbf{g}$ is the gravitational acceleration, $\beta$ is the 
thermal expansion coefficient. $\bm{\sigma_t}$ and $\mathbf{q_t}$ are the turbulent Reynolds stress and the turbulent heat flux given by
$$\bm{\sigma_{t_{ij}}} = -\rho \,  \overline{\statevecU'_i \statevecU'_j} \quad\quad \mathbf{q_{t_i}} = -\rho \,  c_p  \, \overline{\statevecU'_i \Theta'}$$
where $\overline{\statevecU'}$ and $\overline{\Theta'}$ stand for the temporal mean values of the fluctuating velocity and temperature.
The aim of the following study is to solve the constrained nonlinear optimization problem
\begin{equation}\label{ctrl_NS}
\underset{\ctrlvar}{\min} \ \ \functional \left( y \right)
\hspace*{0.5cm}
\txt{subject to }
\hspace*{0.5cm}
\constraint \left(y, \ctrlvar \right) = 0
\end{equation}
where $\functional$ is the functional describing the cost to minimize, $\constraint$ the non-isothermal Navier-Stokes equations \eqref{EQ.Navier_Stokes} and $y(\ctrlvar)$ the state variable which might be represented for example by the velocity field $\statevecU$ or the temperature $\Theta$. In the present article, since the turbulent mixed-convection flow is strongly influenced by the inlet temperature $\theta$ and velocity $U$, we use them as optimization variables. For a given temperature distribution $\hat{\Theta}$, the goal is to recover the inlet velocity $U$ and inlet temperature $\theta$ that minimize the objective functional
\begin{equation}\label{eq.functionals}
\functional(\Theta) = \int_0^{t_f} \int_{\Omega_{int}} (\Theta- \hat{\Theta})^2 \,dx\,dt
\end{equation}
where $[0,t_f]$ is the time frame of simulation and $\Omega_{int}$ the restricted occupied zone of the spatial domain depicted in figure \ref{fig.Temperature_img}.
Two cases of optimization are studied. The first case consists in maintaining the inlet temperature $\theta$ constant and considering the optimization variable to be the inlet velocity $\delta=U$; and the second case by fixing the inlet velocity $U$ and optimizing on the inlet temperature $\delta=\theta$.
%
It is worth mentioning that one could also think about optimizing on   different  parameters, such as the coordinates or the intensity of a heat source in the domain $\Omega$. But, since a GA strategy is to be used, these parameters can directly be incorporated into the cost functional without inducing any modification in the optimization process.
	\subsection{Standard GA approach}
\begin{figure}[hbtp!]
\centering

\includegraphics[width=\linewidth]{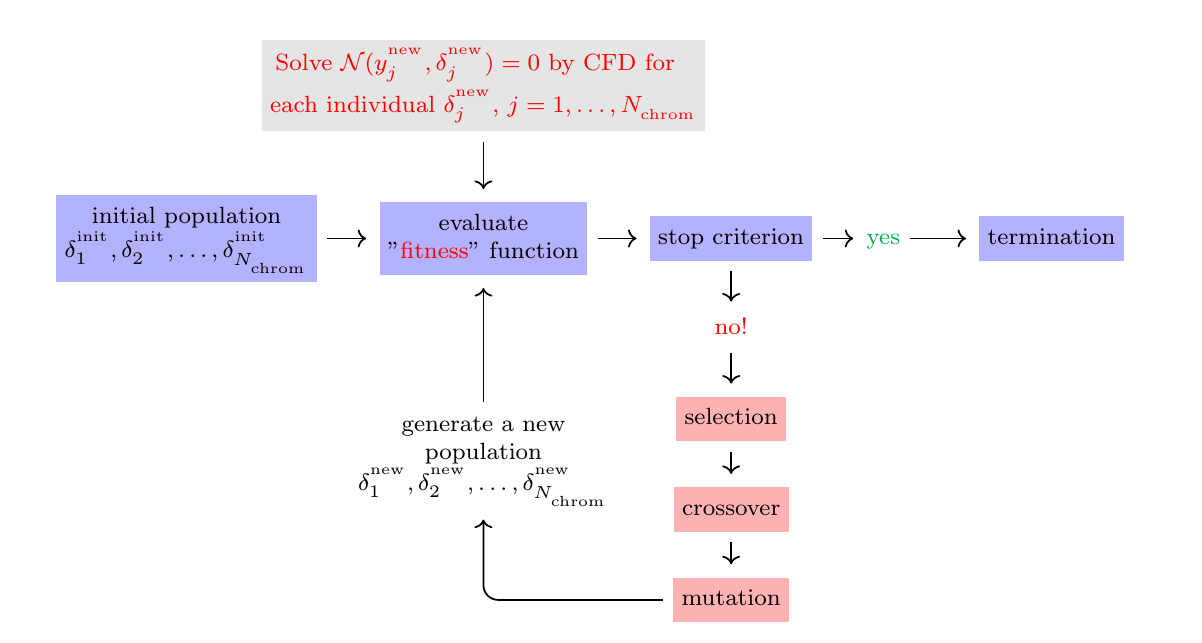}
 \caption{Outline of the Genetic Algorithm.}
\label{flowchart.DDGA}
\end{figure}

		The general idea of GA is illustrated in the flowchart \ref{flowchart.DDGA}. GA consists in starting from a randomly generated set (of size $N_{_{\txt{chrom}}}$) of chromosomes $\ctrlvar_1, \ctrlvar_2, \dots, \ctrlvar_{N_{_{\txt{chrom}}}} $, forming a population. The size of populations is unchanged and fixed to $N_{_{\txt{chrom}}}$. In each population, a fitness value \cite{KOZENY2015} is assigned to each chromosome $\ctrlvar_j$. 
Virtually, any fitness function can be chosen given that no requirement for continuity in the derivatives is needed. Some examples of the choice of fitness functions can be found in \cite{TAMILSELVI2018463, ARORA2017739,ALI201723,VINOTHKUMAR2019191}.
In the present paper, the fitness function $f$ is chosen as the inverse of the objective function, i.e, the fitness of the $j^{\txt{th}}$ chromosome is calculated as follows 
$$f(y_j) = \myfrac{1}{\functional(y_j, {\ctrlvar}_j)} $$
where $y_j$ is obtained by solving the  constraint problem $\constraint(y_j, {\ctrlvar}_j) = 0$.
In order to evolve populations, three main genetic operators \cite{KHOO2002} modeled on the Darwinian concepts of natural selection and evolution are used. These are :
\paragraph{Selection}
There are several methods for selecting the best chromosomes and their transfer to the next generation. In general, a new population of chromosomes is chosen to survive based on their fitness values. That means that a chromosome ${\ctrlvar}_j$ with a large fitness value has higher probability of being reproduced and passed down into the next generation. The probability of reproduction can be calculated as follows
\begin{equation*}\label{selection_probability_formula}
P_s^j =  \left. f(y_j) \middle/ \somme{i}{1}{N_{_{\txt{chrom}}}} f(y_i) \right.
\end{equation*}
Using this reproduction probability, $N_{_{\txt{chrom}}}$ solutions from the current generation are selected by the roulette rule \cite{Goldberg1989} to survive for the next generation. These reproduced solutions are afterwards modulated by the crossover and mutation operators \cite{KHOO2002} . 
\paragraph{Crossover}
The crossover is the operation wherein genes are exchanged between two chromosomes. In particular, all the surviving chromosomes by the roulette selection rule are randomly paired. Precisely, two individuals are randomly selected as parent individuals, then arbitrary positions on both individuals are chosen for crossing locations where exchange of genes takes place. 
In practice, a random number ranging from $0$ to $1$ is generated. If the random number is greater than $P_c$, the two chromosomes in the original pair remain into the next generation. Otherwise, the crossover takes place, and two new chromosomes are created to replace the parent chromosomes. 
%
%
\paragraph{Mutation}
The mutation operator is responsible for bringing new information to the population. With a probability $P_m$ ranging from $0$ to $1$, the mutation operator  accidentally changes one of the resulted genes. 
\\
\\
The above genetic operations are repeated for a predetermined number of generations arbitrarily set by the user.  
The best chromosome of the final generation is declared as the global optimized solution. 
Despite their superiority with respect to other optimization approaches, a serious weakness of high fidelity based GAs is their considerable requirements in computational effort and memory storage \cite{YANG2005}. In fact, GA needs to perform high fidelity simulations many times for each evolved population (iteration). With the increase in the number of generations, the populations and their required crossovers and mutations will increase. These, in turn increase the time complexity of GA, making unfeasible their application in near-real time. In order to tackle this issue, an interpolation strategy suited for non-linear parameterized data and intended to replace the high fidelity solver in the GA is proposed in the next section.
\section{Barycentric interpolation for nonlinear parametrized  data}
	\subsection{Data compression strategy by using the POD}
		Consider a set of parametrized matrices $\{ \trainedSnapMat{k}\in\mbb{R}^{\dimVecorsSnap\times \nbrSnap}, k= 1 ,\dots, \nbrParam\}$ formed from the discrete solutions\footnote{In our case, the solutions $y$ correspond to the turbulent mixed convection temperature distribution $\Theta$, and the parameter $\ctrlvar$ to the inlet velocity $U$ or inlet temperature $\theta$.} $y(\trainingparam{k})$ of a transient non-linear flow problem. i.e,
\begin{equation*}
\trainedSnapMat{k} = \begin{bmatrix}
y(t_{_1}, x_1, \trainingparam{k}) & y(t_{_2}, x_1, \trainingparam{k}) &  \\
&&\\
\vdots & \ddots &  \\
&&\\
y(t_{_1}, x_{\dimVecorsSnap}, \trainingparam{k}) &  & y(t_{_{\nbrSnap}}, x_{\dimVecorsSnap}, \trainingparam{k})
\end{bmatrix}
\end{equation*}
In practice, $\trainingparam{}$ refers to a parameter of the flow problem, $\dimVecorsSnap$ the number of spatial degrees of freedom and $\nbrSnap$ the number of time steps, where it is assumed that $\dimVecorsSnap$ exceeds $\nbrSnap$ by several orders of magnitude. The aim of the following is to extract a set of reduced matrices $\trainedredSnapMat{k}$ that describes the dynamics of the full order matrices $\trainedSnapMat{k}$. To this end, assume that each matrix $\trainedSnapMat{k}$ is approximated in a POD basis \footnote{The POD bases are constructed such that they verify optimality with respect to the Euclidean inner product. In this case, the POD is nothing but the Singular Value Decomposition (SVD). However, other inner products such as $L^2$ or $H^1$ can be used. More details about the POD approach can be found in \cite{Sirovich}.} of dimension $\nbrModes$ as follows
\begin{equation}\label{POD approx sampling}
\trainedSnapMat{k} \approx \Sbase{k} \Tbase{k}^T
\end{equation}
where $\Sbase{i}\in \mbb{R}^{\dimVecorsSnap\times \nbrModes}$ and $\Tbase{i} \in \mbb{R}^{\nbrSnap\times \nbrModes}$ are respectively the spatial and temporal bases.
Now, consider the POD respectively of orders $r$ and $s$, $r,s \leq \nbrModes \nbrParam$, of the column block matrices 
\begin{equation*}
\begin{bmatrix} \Sbase{1} &  \Sbase{2} & \cdots & \Sbase{\nbrBases} \end{bmatrix} = \bm{\Phi}  \Mat{\varphi}^T
\ \ \ \ \ \txt{and} \ \ \ \ \ 
\begin{bmatrix} \Tbase{1} & \Tbase{2} & \cdots & \Tbase{\nbrBases} \end{bmatrix} = \bm{\Lambda}  \Mat{\alpha}^T
\end{equation*}
where $\bm{\Phi}\in \mbb{R}^{\dimVecorsSnap \times r}$, $\Mat{\varphi}\in \mbb{R}^{\nbrModes \nbrBases \times r}$, $\bm{\Lambda}\in \mbb{R}^{\nbrSnap \times s}$, $\Mat{\alpha}\in \mbb{R}^{\nbrModes \nbrBases \times s}$. Let $\varphi_i\in \mbb{R}^{\nbrModes \times r}$ and $\alpha_i\in \mbb{R}^{\nbrModes \times s}$, $i = 1 ,\dots, \nbrParam$, be the column block matrices of $\Mat{\varphi}^T$ and $\Mat{\alpha}^T$ such as
\begin{equation*}
\Mat{\varphi}^T = \begin{bmatrix} \varphi_1 & \varphi_2 & \cdots & \varphi_{\nbrBases} \end{bmatrix} \ \ \ \ \txt{and} \ \ \ \ 
\Mat{\alpha}^T = \begin{bmatrix} \alpha_1 & \alpha_2 & \cdots & \alpha_{\nbrBases} \end{bmatrix}
\end{equation*}
%
It yields that the full order snapshots matrix associated to the parameter $\ctrlvar_k$ can be written as 
\begin{equation}\label{HNIMR snapshots matrix}
\trainedSnapMat{k} \approx\bm{\Phi} \varphi_k \alpha_k^T  \bm{\Lambda}^T
\end{equation}
It is important to note that in the above expression, the change with respect to parameter $\ctrlvar_k$ occurs only on the nested matrices $\trainedredSnapMat{k} =  \varphi_k \beta_k^T$ of significantly reduced size $r\times s$, $r,s\ll N_x$. 
In parametric studies such as optimization, rather than using the full order matrices $ \trainedSnapMat{k}$, it is more convenient to manipulate the corresponding nested reduced matrices $\trainedredSnapMat{k}$ in order to achieve low cost calculations. The interpolation strategy of the matrices $\trainedredSnapMat{k}$ is detailed in the next subsection.
%

	\subsection{Data interpolation}
		In the following, the interpolation approach is first presented for two data samples. The generalization to an arbitrary number of data samples is given afterwards. Let $\trainedredSnapMat{1}$ and $\trainedredSnapMat{2}$ be two parametrized compressed matrices associated respectively to $\ctrlvar_1$ and $\ctrlvar_2$, such that 

$$\trainedredSnapMat{1} =  \varphi_1 \alpha_1^T
\quad\quad
\trainedredSnapMat{2} =  \varphi_2 \alpha_2^T
$$
where $\varphi_k$ and $\alpha_k$ are rank-$q$ parameterized matrices resulted from the data compression procedure. By using the above representations, the goal is to predict the matrix $\untrainedredSnapMat$ associated to a new parameter value $\tilde{\ctrlvar}$ different from $\ctrlvar_1$ and $\ctrlvar_2$. To this end, the barycentric interpolation proposed in \cite{OulghelouSPsDROMArxiv} for subspaces interpolation is used. For the sake of simplicity, we restrict  ourselves to the univariate case and use Lagrange functions to generate interpolation weights. The Lagrange functions constructed by using two points $\ctrlvar_1$ and $\ctrlvar_2$ are given by
$$\omega_1(\tilde{\ctrlvar}) = \myfrac{\tilde{\ctrlvar}-\ctrlvar_2}{\ctrlvar_1-\ctrlvar_2} 
\quad \quad 
\omega_2(\tilde{\ctrlvar}) = \myfrac{\tilde{\ctrlvar}-\ctrlvar_1}{\ctrlvar_2-\ctrlvar_1}$$
During the interpolation process, two sorts of subspaces have to be distinguished. The spatial subspaces $span(\varphi_1)$ and $span(\varphi_2)$, and the temporal subspaces $span(\alpha_1)$ and $span(\alpha_2)$. 
The proposed data interpolation technique suggests to predict the new matrix $\untrainedredSnapMat$ by applying the barycentric interpolation strategy to the spatial and temporal subspaces separately, i.e, it consists in solving the fixed point problems
\begin{equation*}
(\mcal{P}_x)\quad
\begin{cases}
\txt{Find } \tilde{\varphi}\txt{ such that :}
\\
\tilde{\varphi}^T {\varphi}_1 \overset{\tiny\txt{SVD}}{=}\xi_{1} \Sigma_1 \eta_{1}^T
\quad\txt{and}\quad
\tilde{\varphi}^T {\varphi}_2 \overset{\tiny\txt{SVD}}{=}\xi_{2} \Sigma_2 \eta_{2}^T
\\
\tilde{\varphi} = \myfrac{\tilde{\ctrlvar}-\ctrlvar_2}{\ctrlvar_1-\ctrlvar_2} \varphi_1 \tilde{Q}_1 + \myfrac{\tilde{\ctrlvar}-\ctrlvar_1}{\ctrlvar_2-\ctrlvar_1} \varphi_2 \tilde{Q}_2 \quad \txt{where}\quad \tilde{Q}_1 = \eta_{1} \xi_{1}^T \quad\txt{and}\quad \tilde{Q}_2 = \eta_{2} \xi_{2}^T
\end{cases}
\end{equation*}

\begin{equation*}
(\mcal{P}_t)\quad
\begin{cases}
\txt{Find } \tilde{\alpha}\txt{ such that :}
\\
\tilde{\alpha}_1^T {\alpha}_1 \overset{\tiny\txt{SVD}}{=}\zeta_{1} \Upsilon_1 \tau_{1}^T
\quad\txt{and}\quad
\tilde{\alpha}^T \alpha_2\overset{\tiny\txt{SVD}}{=}\zeta_{2} \Upsilon_2 \tau_{2}^T
\\
\tilde{\alpha} = \myfrac{\tilde{\ctrlvar}-\ctrlvar_2}{\ctrlvar_1-\ctrlvar_2} \alpha_1 \tilde{K}_1 + \myfrac{\tilde{\ctrlvar}-\ctrlvar_1}{\ctrlvar_2-\ctrlvar_1} \alpha_2 \tilde{K}_2 \quad \txt{where}\quad \tilde{K}_1 = \tau_{1} \zeta_{1}^T \quad\txt{and}\quad \tilde{K}_2 = \tau_{2} \zeta_{2}^T
\end{cases}
\end{equation*}
The iterative process to solve the problem $(\mcal{P}_x)$ is described by the following fixed point sequence 
\begin{equation*}
(\mcal{P}_x)\quad
\begin{cases}
\tilde{\varphi}^{(0)} \txt{ given, for } n\geq 0
\\
\txt{Perform the SVD of  } \tilde{\varphi}^{(n)^T} {\varphi}_1 \overset{\tiny\txt{SVD}}{=}\xi^{(n)}_{1} \Sigma^{(n)}_1 \eta_{1}^{(n)^T}
\quad\txt{then set }\quad \tilde{Q}_1^{(n)} = \eta_{1}^{(n)} \xi_{1}^{(n)^T} 
\\
\txt{Perform the SVD of  } \tilde{\varphi}^{(n)^T} \varphi_2 \overset{\tiny\txt{SVD}}{=}\xi_{2}^{(n)} \Sigma_2^{(n)} \eta_{2}^{(n)^T} \quad\txt{then set }\quad  \tilde{Q}_2^{(n)} = \eta_{2}^{(n)} \xi_{2}^{(n)^T}
\\
\txt{Update the interpolant as   } \tilde{\varphi}^{(n+1)} = \myfrac{\tilde{\ctrlvar}-\ctrlvar_2}{\ctrlvar_1-\ctrlvar_2} \varphi_1 \tilde{Q}_1^{(n)} + \myfrac{\tilde{\ctrlvar}-\ctrlvar_1}{\ctrlvar_2-\ctrlvar_1} \varphi_2 \tilde{Q}_2^{(n)}
\end{cases}
\end{equation*}
The same strategy applies for the resolution of problem $(\mcal{P}_t)$.
%
%
%
%
%
Now, once the solutions $\tilde{\varphi}$ and $\tilde{\alpha}$ respectively, of the fixed points problems $(\mcal{P}_x)$ and $(\mcal{P}_t)$ are found, the reduced snapshot matrix $\untrainedredSnapMat$ can be formed as
\begin{align*}
\untrainedredSnapMat = \tilde{\varphi} \, \tilde{\alpha}^T =&   \left(\myfrac{\tilde{\ctrlvar}-\ctrlvar_2}{\ctrlvar_1-\ctrlvar_2}\right)^2 \varphi_1 \tilde{Q}_1 \tilde{K}_1^T \alpha_1^T+ \left(\myfrac{\tilde{\ctrlvar}-\ctrlvar_1}{\ctrlvar_2-\ctrlvar_1}\right)^2 \varphi_2 \tilde{Q}_2 \tilde{K}_2^T\alpha_2^T
\\
&+\myfrac{(\tilde{\ctrlvar}-\ctrlvar_1)(\ctrlvar_2-\tilde{\ctrlvar})}{(\ctrlvar_1-\ctrlvar_2)^2}
 \left( \varphi_1 \tilde{Q}_1 \tilde{K}_2^T \alpha_2^T+ \varphi_2 \tilde{Q}_2 \tilde{K}_1^T\alpha_1^T \right)
\end{align*}
A very interesting property of the above formula is that even though space and time reduced bases $\{ \varphi_1, \varphi_2 \}$ and $\{ \alpha_1 , \alpha_2\}$ are separately interpolated, the calibration between the columns of $\tilde{\varphi}$ and $\tilde{\alpha}$ is naturally ensured by the barycentric interpolation, unlike the Bi-CITSGM \cite{oulghelou2020} where the calibration is lost by the Grassmannian interpolation. 
\vspace*{0.2cm}
\\
Let's now state the general framework of the data interpolation approach. To do so, consider a set of parametrized data matrices $\trainedredSnapMat{1}, \cdots, \trainedredSnapMat{\nbrParam}$ associated to the parameter values $\ctrlvar_1, \ctrlvar_2, \dots, \ctrlvar_{\nbrParam}$, such that
$$\trainedredSnapMat{k} =  \varphi_k \alpha_k^T, \quad k=1\dots,\nbrParam$$
The approximate matrix $\untrainedredSnapMat$ for a new untrained value $\tilde{\ctrlvar}\neq \ctrlvar_k$ obtained by solving the following fixed point problems
\begin{equation*}
\hspace*{-1cm}
(\mcal{P}_x)\quad
\begin{cases}
\txt{Find } \tilde{\varphi}\txt{ such that :}
\\
\tilde{\varphi}^T {\varphi}_k \overset{\tiny\txt{SVD}}{=}\xi_{k} \Sigma_k \eta_{k}^T,
\quad
k=1,\dots,\nbrParam
\\
\tilde{\varphi} = \somme{k}{1}{\nbrParam}  \, \omega_k(\tilde{\ctrlvar}) \, \varphi_k \tilde{Q}_k \quad \txt{where}\quad \tilde{Q}_k = \eta_{k} \xi_{k}^T 
\end{cases}
\hfill
(\mcal{P}_t)\quad
\begin{cases}
\txt{Find } \tilde{\alpha}\txt{ such that :}
\\
\tilde{\alpha}^T {\alpha}_k \overset{\tiny\txt{SVD}}{=}\zeta_{k} \Upsilon_k \tau_{k}^T
\quad k=1,\dots,\nbrParam
\\
\tilde{\alpha} = \somme{k}{1}{\nbrParam} \, \kappa_k(\tilde{\ctrlvar}) \, \alpha_k \, \tilde{K}_k \quad \txt{where}\quad \tilde{K}_k = \tau_{k} \zeta_{k}^T
\end{cases}
\end{equation*}
The solution is then constructed as follows 
$$
\untrainedredSnapMat = \somme{k,h}{1}{\nbrParam}  \, \omega_k(\tilde{\ctrlvar}) \, \kappa_h(\tilde{\ctrlvar}) \, \varphi_k \tilde{Q}_k \tilde{K}_h^T\alpha_h^T 
$$
where $\tilde{Q}_k$ and $\tilde{K}_h$ are orthogonal matrices and $\omega_k$ and $\kappa_h$ are some interpolation functions of sum equal to $1$, verifying $\omega_k(\ctrlvar_i) = \kappa_k(\ctrlvar_i) = \bm{\delta^{ki}}$, with $\bm{\delta^{ki}}$ the piecewise Kronecker delta function which value is $1$ if $k$ equals $i$ and $0$ otherwise. The interpolation procedure of nonlinear parametrized data is summarized in algorithm \ref{Alg:barycenter_SPsD}. 
\vspace*{0.2cm}
\\
In order to tackle the severe computational effort of Genetic algorithms, an optimization procedure is proposed in the next section, where algorithm \ref{Alg:barycenter_SPsD} is used as solution predictor instead of the high fidelity solver.
\vspace*{0.2cm}
\\
\begin{algorithm}[H]
\begin{itemize}
\item[]
\begin{itemize}
\item[\textit{\textbf{Offline :}}] 
\end{itemize}
\end{itemize}
Use the POD to compress the trained parametrized data matrices $\trainedSnapMat{k}$ such as
$$ \trainedSnapMat{k} \approx\bm{\Phi} \trainedredSnapMat{k}  \bm{\Lambda}^T \quad\quad \txt{where} \quad \trainedredSnapMat{k} =  \varphi_k \alpha_k^T$$
\begin{itemize}
\item[]
\begin{itemize}
\item[\textit{\textbf{Online :}}]
\end{itemize}
\end{itemize}
Give a value of $\tilde{\ctrlvar}$ (chosen by the user) and calculate the weights $\omega_k(\tilde{\ctrlvar})$ and $\kappa_h(\tilde{\ctrlvar})$ \\
Set $\untrainedredSnapMat^{(0)}  = \tilde{\varphi}_k^{(0)} \tilde{\alpha_k}^{(0)^T}$ arbitrary, for example choose a point $\trainedredSnapMat{k}$ from the sampling\\
\While{ $Error > \varepsilon$ }{
\For{$k\in\{1,\dots,\nbrParam\}$}{
Calculate the matrix $\tilde{Q}_k^{(n)} = \eta_{k}^{(n)} \xi_{k}^{(n)^T}$ where $\tilde{\varphi}^{(n)^T} \varphi_k \overset{\tiny{\txt{SVD}}}{=} \xi_k^{(n)} \Sigma_k^{(n)} \eta_k^{(n)^T}$
\\
Calculate the matrix $\tilde{K}_k^{(n)} = \tau_k^{(n)} \zeta_k^{(n)^T}$  where $\tilde{\alpha}^{(n)^T} \alpha_k \overset{\tiny{\txt{SVD}}}{=} \zeta_k^{(n)} \Upsilon_k^{(n)} \tau_k^{(n)^T}$
}
Update the reduced matrix : $\untrainedredSnapMat^{(n+1)}  =  \somme{k,h}{1}{\nbrParam}  \, \omega_k(\tilde{\ctrlvar}) \, \kappa_h(\tilde{\ctrlvar}) \, \varphi_k \tilde{Q}_k^{(n)} \tilde{K}_h^{(n)^T}\alpha_h^{T}  $\\
Evaluate the error : $Error = \somme{k}{1}{\nbrParam} \somme{h}{1}{\nbrParam} ||\tilde{Q}_k^{(n)} \tilde{K}_h^{(n)^T} - \tilde{Q}_{k}^{(n-1)} \tilde{K}_{h}^{(n-1)^T}||_F$ where $||\cdot||$ denotes the Frobenius norm.
}
\caption{Non-linear data interpolation strategy}
\label{Alg:barycenter_SPsD}
\end{algorithm}

\section{Data-Driven Reduced Genetic Algorithm}
	Basically, the proposed DDGA is a genetic algorithm strategy to solve inverse problems by means of available precomputed parametrized flow data. 
The major advantage of this approach is that the relationship between the state variable $y$ and the optimization variable $\ctrlvar$, earlier established through the mapping $\mcal{N}$, is now replaced by the cheap explicit formula of the barycentric interpolation
\begin{equation}\label{Eq.formula_sol_SPsD}
y(t_l, x_j, \tilde{\ctrlvar}) \approx\bm{\Phi}(x_j)  \untrainedredSnapMat \bm{\Lambda}^T(t_l)
\end{equation}
where $\bm{\Phi}(x_j)$ and $\bm{\Lambda}(t_l)$ denote respectively the  $j^{th}$ and $l^{th}$ rows of the matrices $\bm{\Phi}$ and $  \bm{\Lambda}$, and $\untrainedredSnapMat$ the reduced snapshots matrix to be found by algorithm \ref{Alg:barycenter_SPsD}. 
\vspace*{0.2cm}
\\
In order to make sure that DDGA performs in an optimal manner, the chromosomes are enriched by virtual genes. These genes are the order of POD truncation $q$ and the number of spatial and temporal interpolation neighbors denoted respectively $ne_x$ and $ne_t$. To illustrate this, let $\trainedredSnapMat{1}, \dots, \trainedredSnapMat{4}$ be four reduced matrices associated to the parameter values $\ctrlvar_1 < \ctrlvar_2 < \ctrlvar_3 < \ctrlvar_4 $ respectively such that
$$\trainedredSnapMat{k} = \varphi_k \alpha_k^T,\quad k=1,\dots,4$$
where $\varphi_k$ and $\alpha_k$ are rank-$q$ matrices. Suppose that we want to find an approximation of the reduced matrix $\untrainedredSnapMat$ for an untrained value $\tilde{\ctrlvar}\in ]\ctrlvar_1, \ctrlvar_2[$ by using an order of POD truncation $m < \nbrModes$, three neighbors for spatial interpolation ($ne_x=3$) and two neighbors for temporal interpolation ($ne_t=2$). Then the untrained reduced matrix is approximated as
$$\untrainedredSnapMat  =  \somme{k}{1}{3}  \somme{h}{1}{2} \, \omega_k(\tilde{\ctrlvar}) \, \kappa_h(\tilde{\ctrlvar}) \, \varphi_k \tilde{Q}_k \tilde{K}_h^T\beta_h^T  $$
where the columns of $\varphi_k$ and $\beta_h$ are truncated up to  the order $m$ and 
$$ \omega_k(\tilde{\ctrlvar}) = \displaystyle{\overset{3}{\underset{\underset{i\neq k}{i=1}}{\prod}} \myfrac{\tilde{\ctrlvar}-\ctrlvar_i}{\ctrlvar_k-\ctrlvar_i}}
\quad\quad \txt{and} \quad \quad
\kappa_h(\tilde{\ctrlvar}) = \displaystyle{\overset{2}{\underset{\underset{i\neq h}{i=1}}{\prod}} \myfrac{\tilde{\ctrlvar}-\ctrlvar_i}{\ctrlvar_h-\ctrlvar_i}}
$$
In the proposed genetic algorithm strategy, the $j^{\txt{th}}$ chromosome is then the candidate $\bar{\ctrlvar}_j = \{\ctrlvar_j, ne_t, ne_x, m \}$ where  $\ctrlvar_j$, $ne_t$, $ne_x$ and $q$ are its genes. Accordingly, the original optimization problem \eqref{ctrl_NS} is modified yielding to
\begin{equation*}
\underset{\bar{\ctrlvar}}{\min} \ \ \functional(\untrainedredSnapMat , \bar{\ctrlvar} )
\hspace*{0.5cm}
\txt{such that } \untrainedredSnapMat \txt{ is the output of algorithm } \ref{Alg:barycenter_SPsD}
\end{equation*}
In the next section, the potential of this approach is assessed on the inverse parameter identification problem involving a turbulent mixed convection flow.

\section{Numerical experiments}
		In this section, the CFD model used to solve the mixed-convection problem is first validated with respect to the benchmark experimental data. Then a set of solutions sampled in different time instants and different trained parameters are created and eventually used to assess the efficiency of the proposed  DDGA. 
\subsection{CFD validation}
\begin{figure}[h!]
\begin{subfigure}{0.5\linewidth}
\includegraphics[width=\linewidth]{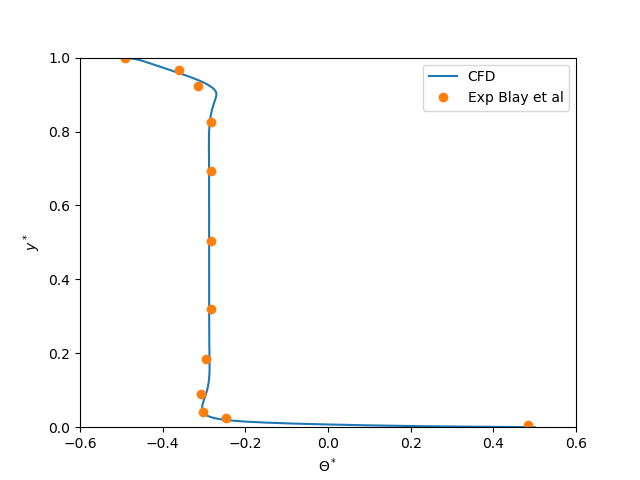}
\label{comp_temperature_a}
\caption{$\Theta^*$ at $x^*=0.5$}
\end{subfigure}%
\begin{subfigure}{0.5\linewidth}
\includegraphics[width=\linewidth]{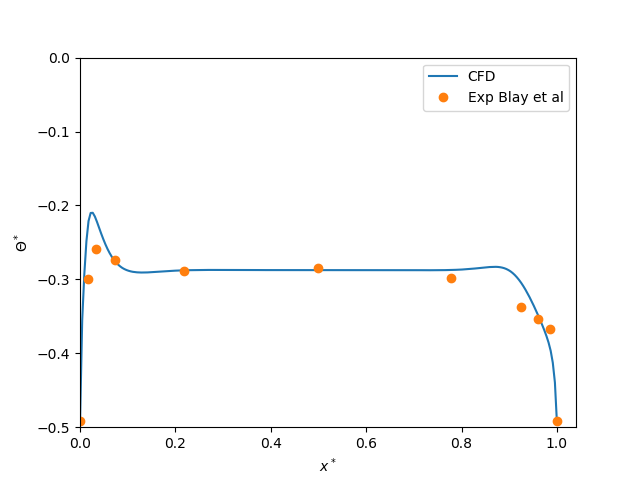}
\label{comp_temperature_b}
\caption{$\Theta^*$ at $x^*=0.5$}
\end{subfigure}
\caption{Comparison between the numerical results (CFD) and the experimental results (Exp Blay et al.) of $\Theta^*$ at $x^*=0.5$ (a) and 
       $y^*=0.5$ (b)}
\label{comp_temperature}
\end{figure}

This series of numerical computations was based on the experiment carried out by Blay et al. \cite{Blay_1992}, where a turbulent mixed convection flow was generated
in a ventilated cavity with dimensions $1.04\times1.04\times0.7$ $m^3$. In this experiment, a two-dimensional flow was generated in the enclosure shown in figure \ref{fig.Temperature_img}, which was surrounded by two guard cavities. 
The reference temperature $\Theta_0$ was the average temperature in the cavity. The Rayleigh number 
of this configuration, based on the cavity height and on the temperature
difference between the heated floor ($\theta_{hot}=35.5^oC$) and the other walls and the inlet ($\theta_{cold}=\theta=15^oC$), was $2.13\times 10^9$. The Reynolds number based on 
the air velocity at inlet $U=0.57m/s$ and on the inlet height was 654. The two-dimensional turbulent flow was
modeled with the RNG k-epsilon model \cite{Yakhot_1992}. To compute this flow, and to generate all input data necessary for the study presented in this paper,
the finite volume code OpenFOAM \cite{OpenFOAM} was used. The computational domain was discretized into a non uniform grid made of $120000$ hexaedral cells,
which was very tight close to the walls, in order to properly discretize the boundary layer. The non-isothermal flow described by equations \eqref{EQ.Navier_Stokes} was calculated with the buoyantBoussinesqPimpleFoam solver. At the inlet, the velocity boundary conditions were $u=0.57$ m/s and $v=0$ m/s,
the temperature was $\Theta=15^oC$, and the turbulent boundary conditions were $k=1.25\times 10^{-3}m^2/s^2$ and $\epsilon=5.76\times 10^{-3}m^2/s^3$. On the
walls, no-slip boundary conditions were applied for the velocity components, the temperature was equal to $35^oC$ on the floor, and to $15^oC
$ on the other walls. At the outlet, zero gradient boundary conditions were applied for the temperature, the velocity components and the turbulent variables. 
The steady flow presented in this paragraph was reached by computing an unsteady flow, starting at $t=0$ s from $\Theta_{ini}=\theta_{cold}$ for the temperature,
and $u_{ini}=v_{ini}=0$ for the velocity components.
The convection terms were discretized with the Gauss linear Upwind scheme,
and the laplacian terms were approximated with the Gauss linear corrected scheme. With this non uniform mesh, the average $y^+$ value was equal to 1.1, and 
the maximum value was $3.3$. In figure \ref{comp_temperature}, the temperature profiles at $x=0.52$ and at $y=0.52$ are shown such that, $\Theta^*=\frac{\Theta-\Theta_0}{\theta_{hot}-\theta_{cold}}$, $x^*=x/H$ and $y^*=y/H$ where $H$ is the cavity height. A satisfactory agreement can be noticed.

\subsection{Optimization problem of the mixed convection flow}
\begin{figure}[h!]
\includegraphics[width=\linewidth]{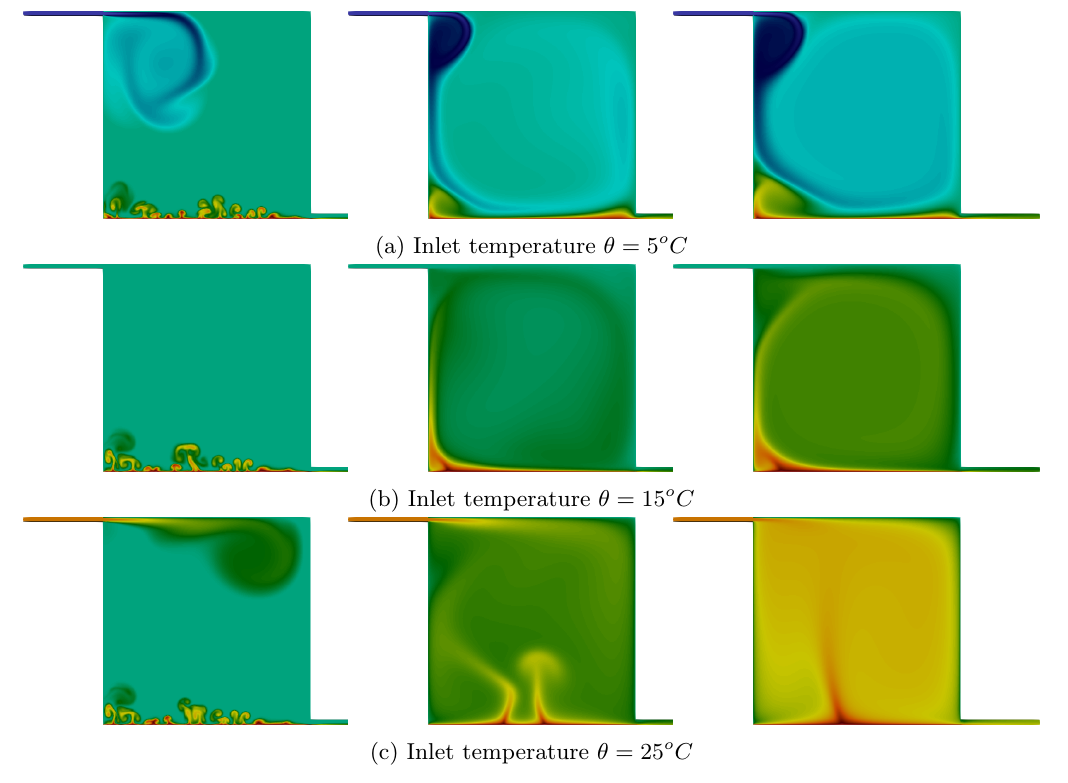}
\caption{Temperature distribution at three time instants $t=8.75s$ (left), $t=55s$ (middle) and $t=1250s$ (right) for the case of variable inlet temperature}
\label{fig.3temperature_snapshots_Tinj}
\end{figure}

\begin{figure}[h!]
\includegraphics[width=\linewidth]{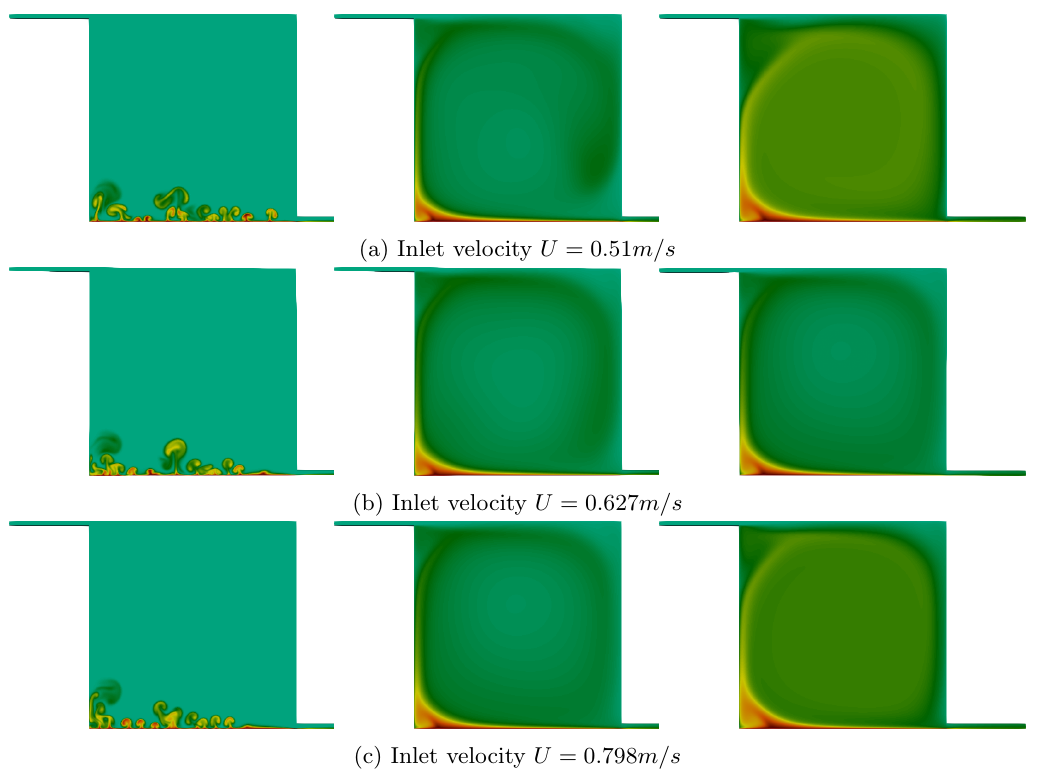}
\caption{Temperature distribution at three time instants $t=8.75s$ (left), $t=55s$ (middle) and $t=1250s$ (right) for the case of variable inlet velocity.}
\label{fig.3temperature_snapshots_Uinj}
\end{figure}

As claimed in the earlier section, a data driven approach (algorithm \ref{Alg:barycenter_SPsD}) is to be embedded within the GA in order to tackle the severe computational effort due to high fidelity simulations. 
Thereby, the time required
for evaluating the fitness of one chromosome passes from several hours to real time, and thus, drastically reducing the time needed for optimization. 
By using a set of parametrized flow solutions, the goal is to act on  the inlet velocity or temperature in order to minimize the discrete cost functional 
\begin{equation}\label{eq.functionals}
\functional(\Theta) = \myfrac{1}{\nbrSnap} \somme{n}{1}{\nbrSnap}  \int_{\Omega_{int}} (\Theta^n- \hat{\Theta}^n)^2 \,dx
\end{equation}
where the interior subdomain represented in figure \ref{fig.Temperature_img} is considered such that $\Omega_{int} = [0.1, 0.9]\times[0.15,0.7]$. The superscript $n$ refers to the time instant, $\Theta^n$ the calculated temperature and $\hat{\Theta}^n$ the target temperature.
\vspace*{0.2cm}
\\
%
%
%
%
A set of training simulations, based on the configuration presented in figure 1, for different values of inlet velocity $U$ and inlet temperature $\theta$ were performed with OpenFOAM over the time interval $[0,t_f]$. For all the cases considered in this paper, at $t=0$ s, the temperature in the cavity is equal to $\theta_{cold}$, and the velocity to $\boldmath{0}$. The final time instant $t_f$ was chosen in such a way that the temporal evolution of the temperature in the center of the cavity did not vary according to time. For all simulations, $t_f=1250$ s was a sufficiently long time interval.
$1000$ snapshots uniformly spaced in the time interval $[0,t_f]$ are then used to build the temperature POD decompositions, where the maximal POD truncation order $q$ is initially set to $60$. Two series of tests are carried out :
\vspace*{0.3cm}
\\
\noindent\textit{Test Series 1} : the optimization is performed by fixing the inlet temperature $\theta = 15^{\circ} C$ and varying the inlet velocity $U$. The following three values of $U$ are considered for the training phase : $0.51m/s$, $0.627m/s$ and $0.798m/s$.  
Knowing the temperature $\hat{\Theta}$ in the subdomain $\Omega_{int} $, the aim is to determine by applying DDGA, the corresponding inlet velocity $\hat{U}$ with values : $0.54m/s$, $0.57m/s$, $0.5985m/s$, $0.67m/s$, $0.7125m/s$ and $0.755m/s$. Recall that besides the inlet velocity $U$, the space of search by DDGA is enriched by the order of truncation of the POD decompositions $m$, and the number of temporal and spatial neighboring subspaces $ne_t$ and $ne_x$, selected to perform the barycentric interpolation. For this case, the DDGA is allowed to search in the following space
$$ K = \left\{(U, ne_t, ne_x, m)\in \mbb{R}_+ \times\mbb{N}^3, \hspace*{0.3cm} 0.51 \leq U \leq 0.798; \, 2 \leq ne_t, ne_x \leq 3 \hspace*{0.2cm}\txt{and}\hspace*{0.1cm} 4 \leq  m \leq \nbrModes \right\}$$
In order to analyze the performance of the method proposed in this paper, it is interesting to have a look at the isovalues of temperature and velocity magnitude in the cavity for the three training values of inlet velocity (see figures \ref{fig.3temperature_snapshots_Tinj}). At the beginning of all simulations presented in this paper, the air in the vicinity of the hot floor is warmed by thermal diffusion, and it is then lifted by natural convection  along the hot floor (one can notice small thermal plumes at the beginning of all simulations). For an inlet velocity between 0.51 and 0.798 m/s, and an inlet temperature value of $15^oC$, a clockwise recirculation region is generated by the combined effects of the forced convection induced by the air injection, and of the natural convection which occurs along the hot floor. 
\\
\textit{Test Series 2} : in this case, the optimization is performed by acting on the inlet temperature $\theta$ while the inlet velocity is set to the fixed value $0.57 m/s$. The considered training injection temperature values are : $5^{\circ} C$, $10^{\circ} C$, $15^{\circ} C$, $20^{\circ} C$ and $25^{\circ} C$. As in test series 1, the aim is to use DDGA to approximate the optimal inlet temperature $\hat{\theta}$ with values : $7.5^{\circ} C$, $12.5^{\circ} C$, $17.5^{\circ} C$ and $22.5^{\circ} C$ associated to the known temperature distribution $\hat{\Theta}$. The space of search by DDGA in this case is given by
$$ K = \left\{(\theta, ne_t, ne_x, m)\in \mbb{R}_+ \times\mbb{N}^3, \hspace*{0.3cm} 5 \leq \theta \leq 25; \, 2 \leq ne_t, ne_x \leq 5 \hspace*{0.2cm}\txt{and}\hspace*{0.1cm} 4 \leq  m \leq \nbrModes \right\}$$
For this test series, let us have a look at the isovalues of temperature and velocity magnitude
obtained for the inlet temperatures of $5^oC$, $15^oC$ and $25^oC$ (see figures \ref{fig.3temperature_snapshots_Uinj}). For small inlet temperatures
($\theta=5^oC$), the air 
in the upper left part of the cavity, which is too cold, falls along the left wall. It is then warmed by the hot floor, and lifted by natural convection
with a counterclockwise motion along the hot floor. For higher inlet temperatures ($\theta=15^oC$), the air in the upper part of the cavity is warm and the clockwise motion of a large recirculation region induced by the combined effects of the forced convection phenomenon and the natural convection phenomenon along the hot floor can be seen. For the highest temperature velocities ($\theta=25^oC$), the injected air is hot, it remains in a large region along the ceiling, it falls along the left and right cold walls, and is lifted along the heated floor, inducing two recirculation regions, a clockwise one in the right part of the cavity, and a coutnterclockwise one in the left part of the cavity. For this second series of training tests, it can be concluded that for various inlet velocities, the flow regimes are different from each other.
\vspace*{0.2cm}
\\
In the numerical experiments of DDGA, a population of $20$ chromosomes formed by $4$ genes randomly generated in $K$ is used as initial guess to run the DDGA. The algorithm is allowed to run until a maximum number of iterations predetermined by the user is reached. The maximum number of iterations here is set to $30$.

%
%
%

%
%
	\subsection{Temperature optimization by the proposed Data-driven Genetic Algorithm}
\begin{figure}[hbtp!]
\centering
\includegraphics[width=\linewidth]{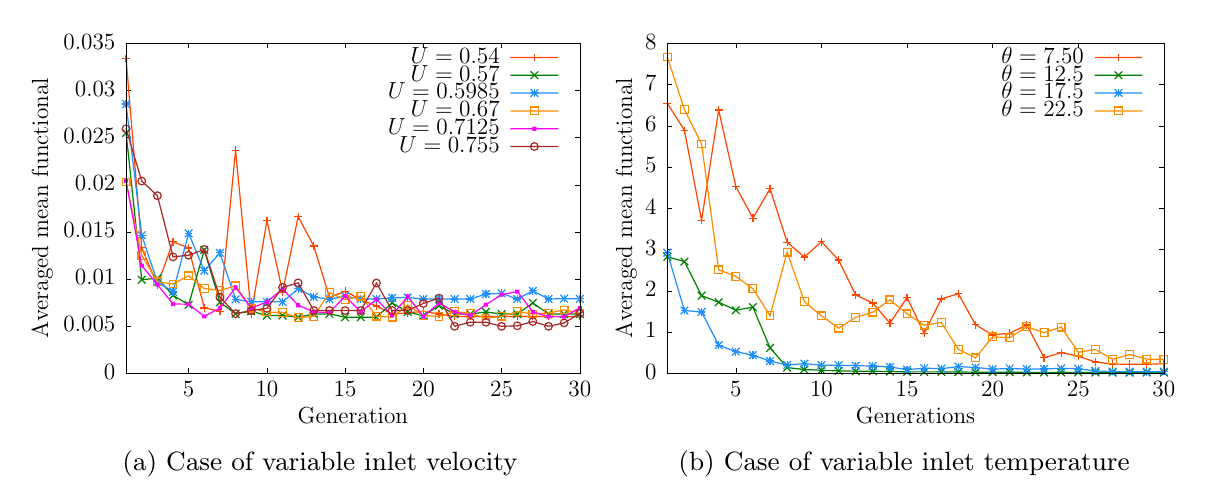}
\caption{Evolution of the averaged functional over generations of DDGA for the cases of variable inlet velocity and variable inlet temperature.}
\label{average_func_cavity}
\end{figure}

\begin{figure}[hbtp!]
\centering
\includegraphics[width=\linewidth]{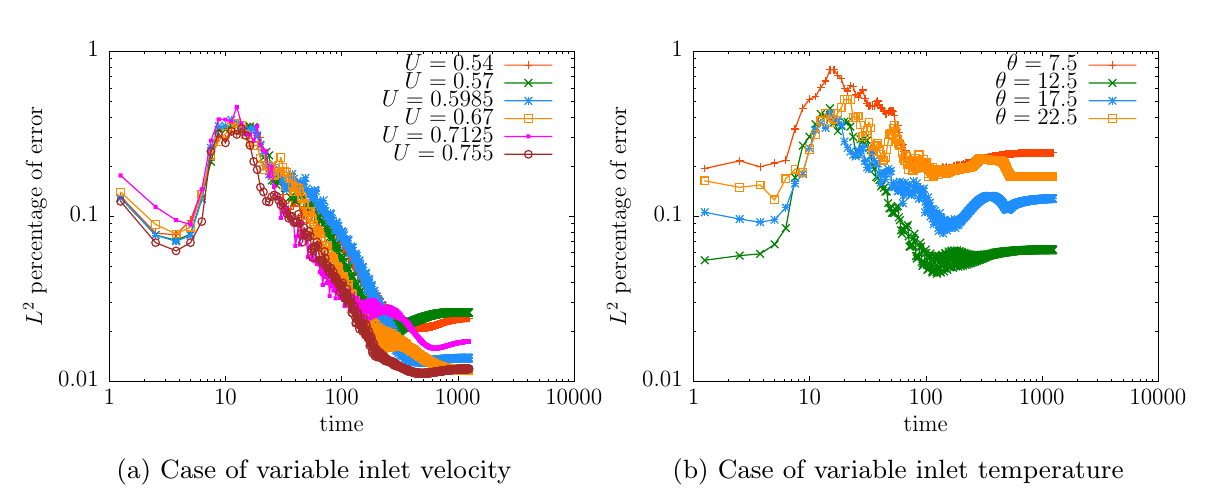}
\caption{Percentage of error of the converged temperature solution by DDGA.}
\label{L2_Errors_Temperature}
\end{figure}

\begin{figure}[hbtp!]
\includegraphics[width=\linewidth]{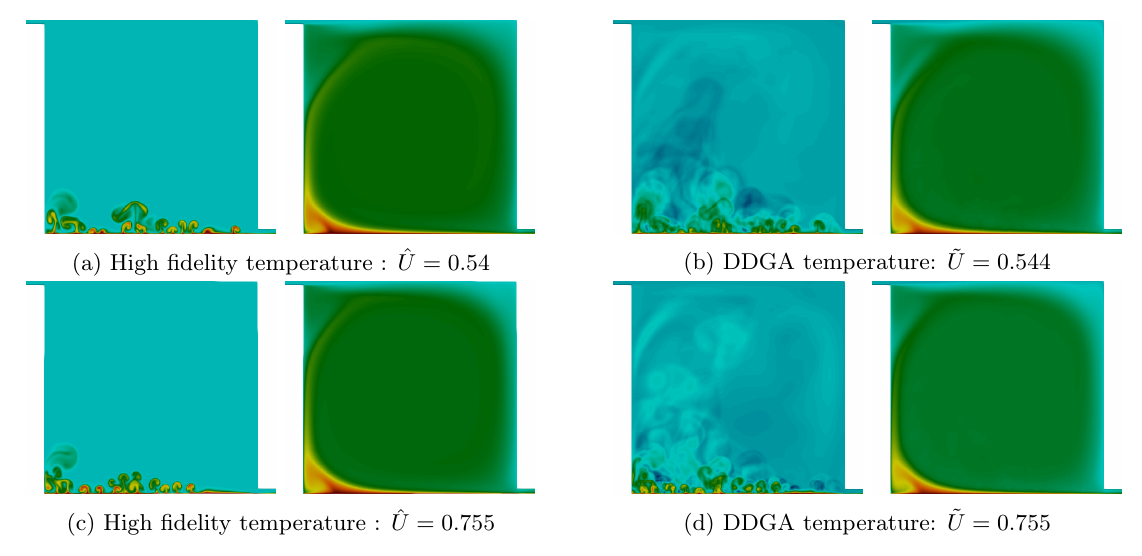}
\caption{Comparison of the high fidelity and DDGA temperature solutions at two different instants of the flow, for the case of variable inlet velocity. The odd columns describe the first appearance of thermal plumes at $t=8.75s$, and the even columns the established regime of the temperature at $t=1250s$. }
\label{fig:Comparison_DDGA_and_optimal_Temp_Vinj}
\end{figure}

\begin{figure}[hbtp!]
\includegraphics[width=\linewidth]{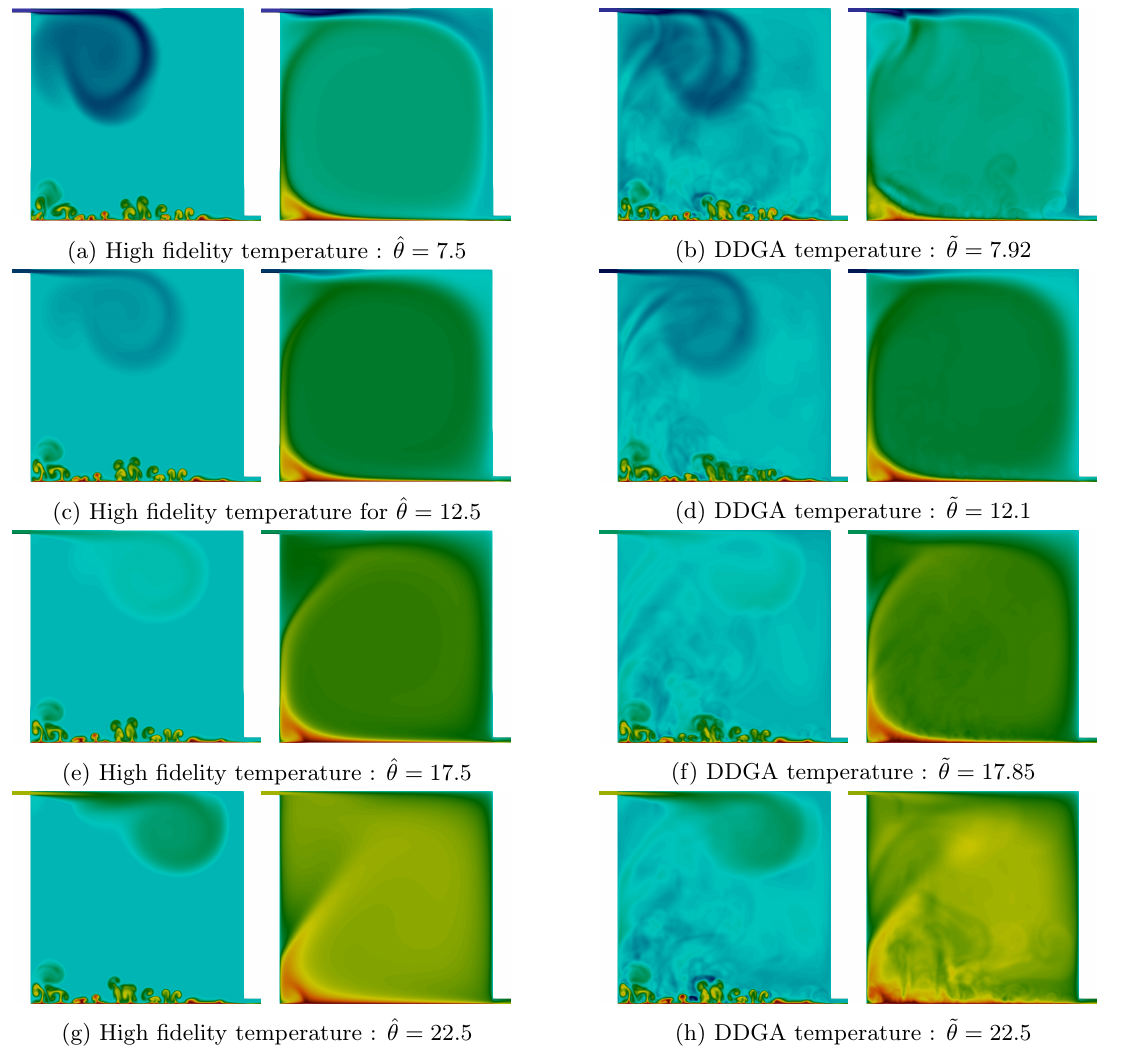}
\caption{Comparison of the high fidelity and DDGA temperature solutions at two different instants of the flow, for the case of variable inlet temperature. The odd columns describe the first appearance of thermal plumes at $t=8.75s$, and the even columns the established regime of temperature at $t=1250s$. }
\label{fig:Comparison_DDGA_and_optimal_T}
\end{figure}

		In the following, the results of the inverse parameter identification problem involving the turbulent mixed convection flow are presented and analyzed. 
The decay of the averaged functional for the cases of variable inlet temperature and variable inlet velocity is plotted in figure \ref{average_func_cavity}. It shows that after successive generations, the averaged cost decreases and tends to stagnate, meaning that the populations contain a chromosome of high recurrence. This chromosome is eventually considered as the best individual that approximates the sought optimum of the inverse problem. The outputs of this best chromosome from the last generation are listed in Table \ref{tab:results_V_inj} and Table \ref{tab:results_T_inj}. It can be seen that the DDGA succeeded to recover approximations $\tilde{U}$ and $\tilde{\theta}$ of the sought optimal inlet values $\hat{U}$ and $\hat{\theta}$ with good accuracies. Moreover, The $L^2$ percentage of error over the simulation time interval between the target temperature and the solution obtained by DDGA for all the cases was less than $0.8\%$ (see figure \ref{L2_Errors_Temperature}). Figures  \ref{fig:Comparison_DDGA_and_optimal_Temp_Vinj} and \ref{fig:Comparison_DDGA_and_optimal_T} show the target temperature solutions side by side with the reconstructed temperature solutions obtained at the end of DDGA. The odd columns show the first appearance of the thermal plumes that emerge from the heated bottom wall of the cavity, while the even columns represent the temperature distribution in its established regime. From a visual perspective, it can be seen that the approached solutions by DDGA are in good agreement with the target high fidelity solutions. The converged DDGA-solution succeeded to track the provided target temperature catching by that the most of the dynamics features present in the temperature along the simulation time interval and all over the domain $\Omega$.
More particularly, for the first test series which led to similar features but different values of velocity and temperature, the velocity 
and temperature values in the cavity are properly recovered by the method proposed in this paper. It can also be pointed out that for the second test 
series which involved various flow regimes
and which was much more complex than the first case, the new method presented here provided results that showed a good accuracy.
Here, the attention of the reader is bounced back to the fact that the POD truncation order $m$ as well as the neighbors number $ne_t$ and $ne_x$, are extremely important parameters of DDGA. These parameters are essentially meant to ensure the good performance of the barycentric interpolation inside the DDGA. By analyzing the results of tables \ref{tab:results_V_inj} and \ref{tab:results_T_inj}, we observe that these quantities vary from a test case to another, i.e, variable neighbors number with less than $12$ modes were needed to represent the DDGA-optimal flow for the case of inlet velocity, while the case of variable inlet temperature has more complicated dynamics and needed at least $25$ modes to represent the solution. This confirms that besides the ability to locate a global optimum of the inlet problem, the DDGA has the feature to eliminate the noise that might intervene from further data samples and from lower frequency POD modes.
Finally, in terms of computational effort, DDGA is very efficient and performs in near-real time. The overall computational time needed to perform $30$ generations in a single cluster was less than two minutes. In inverse problems of turbulent flows, this represents a tremendous gain in CPU time compared to traditionally used high fidelity approaches.


%

\begin{table}[hbtp!]
 \begin{tabular}{ c|ccccc }
Sought optimal values & Approximated value & $ne_t$ & $ne_x$ &trunc. order $m$  \\
 \hline
$\hat{U} = 0.54$ & 	$\tilde{U} = 0.544$	&    $2$   	& $2$ & $8$ \\
$\hat{U} = 0.57$ & $\tilde{U} = 0.578$	&    $2$   	& $2$ & $10$ \\
$\hat{U} = 0.5985$ & $\tilde{U} = 0.59$	&    $2$   	& $2$ & $10$ \\
$\hat{U} = 0.67$ & $\tilde{U} = 0.66$	&    $3$   	& $3$ & $10$ \\
$\hat{U} = 0.7125$ & $\tilde{U} = 0.706$	&    $2$   	& $3$ & $7$ \\
$\hat{U} = 0.755$ & $\tilde{U} = 0.755$	&    $3$   	& $3$ & $12$ \\
\hline
\end{tabular}
\caption{Outputs of the optimal control by using DDGA for the case of variable inlet velocity.}
\label{tab:results_V_inj}
\end{table}
%
%
%
%
\begin{table}[hbtp!]
 \begin{tabular}{ c|ccccc }
Sought optimal values & Approximated value & $ne_t$ & $ne_x$ &trunc. order $m$  \\
 \hline
$\hat{\theta} = 7.50$ & 	$\tilde{\theta} = 7.92$	&    $4$   	& $4$ & $25$ \\
$\hat{\theta} = 12.5$ & $\tilde{\theta} = 12.10$	&    $2$   	& $2$ & $31$ \\
$\hat{\theta} = 17.5$ & $\tilde{\theta} = 17.85$	&    $2$   	& $2$ & $38$ \\
$\hat{\theta} = 22.5$ & $\tilde{\theta} = 22.50$	&    $3$   	& $4$ & $30$ \\
\hline
\end{tabular}
\caption{Outputs of the optimal control by using DDGA for the case of variable inlet temperature.}
\label{tab:results_T_inj}
\end{table}
%
%
%
%
%
%
%
%
%
%
%
%
%
%
%
%
%
%
%
%
%
%
%
%
%
%
%
%

%

\section{Conclusions}
In  this  paper,  we  have  proposed the data driven optimization approach DDGA by combining genetic algorithms and the barycentric interpolation. The barycentric interpolation is presented here as an equation-free approach that allows to learn from trained data solutions and predict the evolution of new untrained solutions without any knowledge of the physics hidden behind. 
The numerical assessments of DDGA are performed on the inverse problem involving a turbulent mixed convection problem, where the variation is carried out on the inlet velocity and then on the inlet temperature. We notice that DDGA succeeded to track the optimal solutions and to deliver satisfying approximations in less than two minutes. This significant gain endorses the great potential of this approach compared to a high fidelity based GA that could last for many hours or days.
\section*{Acknowledgement}
This material is based upon work financially supported by CPER BATIMENT DURABLE - Axe 3 "Qualité des Environnement Intérieurs (QEI)" (P-2017-BAFE-102) and French Astrid ANR MODULO'PI (ANR-16-ASTR-0018 MODUL’O $\Pi$).

%
%

\section*{\large References}
\bibliographystyle{ieeetr}
\bibliography{./BIBLIO}
\end{document}